\documentclass[aps,pra,12pt]{revtex4-1}
\usepackage[colorlinks=true, citecolor=blue,linkcolor=blue, urlcolor=blue]{hyperref}
\usepackage[utf8]{inputenc}
\usepackage{amssymb}
\usepackage{amsthm}
\usepackage{amsmath}
\usepackage{graphicx}
\usepackage{flushend}
\begin{document}
\title{The role of anharmonicity in single-molecule spin-crossover}
\author{Chuan Guan}
\author{Yun-An Yan}
\email{yunan@ldu.edu.cn}
\affiliation{School of Physics and Optoelectronic Engineering, Ludong University,
    Yantai, Shandong 264025, China}
\begin{abstract}
We exploit the system-bath paradigm to investigate vibrational-anharmonicity effects 
on spin-crossover in a single molecule. Focusing on weak coupling, we use the linear response approximation 
to deal with the nonlinear vibrational bath and propagate the Redfield master equation to obtain the equilibrium high 
spin fraction. We take both the anharmonicity in the bath potentials and the nonlinearity 
in the spin-vibration coupling into account and find a strong interplay between these two effects. 
Further, we show that the spin-crossover in a single molecule is always a gradual transition and
the anharmonicity-induced phonon drag greatly affects the transition behavior.
\end{abstract}

\maketitle

\section{Introduction}

Octahedral first-row transition-metal complexes of 3$d^{4} $-3$d^{7}$ 
may switch between the high-spin (HS) and the low-spin (LS) states
under external perturbations, 
such as temperature change~\cite{gutlich94_2024}, 
light irradiation~\cite{hauser1986reversibility}, 
pressure~\cite{molnar2003raman}, 
magnetic field~\cite{bousseksou2002dynamic},
and electric field~\cite{prins2011room}.
Such a spin-crossover (SCO)~\cite{cambi31_2591,brooker2015spin,bousseksou11_3313} 
usually comes up with
changes in magnetic moment, color~\cite{kahn1992spin}, structure, 
dielectric constant~\cite{bousseksou2003observationJ}, 
and even catalytic capacity~\cite{zhong2018c}. Because of the rich physics and phenomena, SCO has wide potential 
applications in molecular switches~\cite{bhandary2021designing}, 
memory devices~\cite{kahn98_44}, sensors~\cite{s120404479}, 
actuators~\cite{molnar2018spin} and has aroused extensive research interests.

Among various SCO phenomena, the temperature-induced ones~\cite{shongwe2007thermally} are of special interest, 
in which energy splitting between the HS and the LS states are close to the thermal energy. 
In practical applications, the transition is required to be abrupt and occur at room temperature.
In this case the vibrations will play 
a key role due to the energy matching~\cite{ronayne2006vibrational,zhang2014predicting}.  
Various models have been proposed to understand the role of vibrations, including 
the Ising model extended with lattice vibration~\cite{ZIMMERMANN1977779}, the atom-phonon model~\cite{nasser2011two}, 
and the stretching-bending model~\cite{ye2015monte}.

Note that in metal-organic compounds, vibrations often assume strong anharmonicity due to 
the presence of hydrogen bonding or other intermolecular interactions. Shelest studied the thermodynamics 
of SCO with an anharmonic model and revealed that anharmonicity is one important parameter controlling 
the SCO transition~\cite{shelest2016role}. Nicolazzi \emph{et al.} used the Lennard-Jones potential to 
model intermolecular interactions and found that anharmonicity can reduce the transition temperature and 
make the HS state more stable~\cite{nicolazzi2008two,nicolazzi2013elastic,mikolasek2017surface}. 
These authors further demonstrated that anharmonicity in intermolecular interactions is pivotal to understand SCO 
in nanostructures which allows atoms to undergo large displacements away from their equilibrium 
positions~\cite{fahs:hal-04113277}. Boukheddaden proposed an anharmonic coupling model 
and showed that change in anharmonicity drastically alters the SCO transition~\cite{boukheddaden2004anharmonic}.  
However, there are controversial conclusions in the literature. For instance, Wu \emph{et al.} 
performed density functional theory calculations for the effects of anharmonicity on the zero-point energy 
and the entropy in Fe(II) and Fe(III) complexes and found a rather small contribution to SCO~\cite{wu2019role}.

Besides in metal-organic compounds, anharmonicity exists in a wide variety of systems and 
has been attracting increasing research interests. 
Even though suppressed in crystals by the crystal symmetry, lattice anharmonicity
can still significantly affect crystal's kinetics, dynamics, and thermodynamics~\cite{cowley63_421,manley19_1928}.
This issue becomes more profound on surfaces~\cite{wertheim94_2277} or in disordered systems, 
including polar liquids~\cite{fleming96_109}, 
glasses~\cite{baldi14_125502}, and
molecular systems~\cite{yan11_5254,galestian17_24752}. 
For example, Lunghi and coworkers found that 
the anharmonicity in single molecule magnets is responsible for fast 
under-barrier spin relaxation~\cite{lunghi17_14620}.
Now the anharmonicity-induced nonlinear effects and the underlying origin 
can be accessed with linear and ultrafast 
IR and Raman spectroscopies thanks to their tremendous progress 
in the past three decades~\cite{passino97_6094,tayagaki01_2886,okumura01_237,fulmer04_8067,ould_hamouda18_385,bec18_483,lada22_108}.

When dealing with anharmonicity, available studies were based on either a classical 
description or the noninteracting, independent quantum oscillator model. 
Here we suggest a consistent quantum 
approach by using nonlinear quantum dissipation.
In order to obtain a clear picture of the anharmonic effect itself and to avoid discussing complicated 
interplays between anharmonicity, spin pairing, energy splitting, and interactions among transition-metal centers, 
we focus on the single-molecule SCO transition in the weak spin-vibration coupling regime.
A numerically exact simulation of an intrinsic nonlinear dissipation system is expensive. 
For the weak dissipation under study, 
we can approximate the bath with the anharmonic influence functional approach suggested by 
Makri \emph{et al.}~\cite{ilk1994real,makri1999linear,makri1999iterative}.  Eventually, the anharmonic influence functional approach 
maps the anharmonic bath to a harmonic one with the help of an effective spectral density function. 
After doing so, we can follow the quantum master equation (QME) approach developed for the linear dissipation to 
investigate single-molecule SCO. Note that simulations of SCO with QME was recently used 
by Orlov \emph{et al.}~\cite{orlov2021light,orlov2022light}.

At first glance the use of a nonlinear bath model is \emph{ad hoc} 
because the bath is always linear in the open system paradigm at any given temperature. 
Note that the dissipation theory is a phenomenological description of open systems 
based on the quantum fluctuation-dissipation theorem~\cite{mori65_423,kubo66_255}.
One of the key points is the system-bath separation 
which implies that the system-bath interaction is weak and 
the effect of a single bath mode on the system is negligibly small. 
The effect of the bath, therefore, is a collective behavior 
and the system only feels the overall environmental fluctuation that follows the 
Gaussian statistics. The fluctuation of the bath is characterized by its spectral density function and
can be reproduced with the linear-dissipation Caldeira-Leggett model~\cite{caldeira81_211}. 
This framework, however, only holds at a fixed temperature and in principle
the spectral density functions at different temperatures are not the same.
In reality, all baths, especially for low-dimensional and molecular systems, are intrinsically anharmonic.  
The use of linear dissipation for simulating 
physics of an anharmonic system will have to adopt a different spectral density function, hence a different bath, 
for each temperature and therefore loses the predictive power for any temperature-dependent behavior. 
A remedy is the above-mentioned nonlinear dissipation model, in which the same bath 
is used to consistently yield the effective spectral density functions for all temperatures.

The rest of the paper are organized as follows. In Sec. II, we outline the anharmonic influence 
functional method in the linear response regime. In Sec. III, we present  
calculations with different anharmonic settings 
to check their effects on SCO. A concise summary and outlook are provided in Sec. IV.

\section{Theoretical model}
We only investigate the one-step SCO transition which is dictated by the Hamiltonian
\begin{align}
\label{eq:Htot1}
\hat{H} _{tot} =\begin{pmatrix}\epsilon_{L}+ \hat{H}_{vib}^{(L)} & -\frac{\lambda}{2} \\ -\frac{\lambda}{2} &\epsilon_{H}+ \hat{H}_{vib}^{(H)} \end{pmatrix},
\end{align}
where $\lambda$ is the spin-orbital coupling, $\epsilon_S$ are the electronic energy, and 
$\hat{H}_{vib}^{(S)}$ with $S=L$ and $H$ describe the vibrations of the LS and the HS states, respectively.
Here we discuss the anharmonic effects without considering the Duschinsky rotation and mode-mode coupling. 
Furthermore, we assume that the LS and the HS states have the same anharmonicity. 
With these approximations and up to the fourth order, the vibrational Hamiltonians 
can be extracted from separate, \emph{ab initio} anharmonic force constant calculations at the LS and the
HS states~\cite{giese2006multidimensional}, yielding
$\hat{H}_{vib}^{(S)}=\sum_{j}\left\{\frac{\hat{p}_{j}^{2} }{2m_{j}}+ v^{(S)}_{j,1} \hat{x}_j 
+ v^{(S)}_{j,2} \hat{x}_j^2 + v^{(S)}_{j,3} \hat{x}_j^3 + v^{(S)}_{j,4} \hat{x}_j^4 \right\}$ with
$v_{j,3}^{(L)} = v_{j,3}^{(H)}$ and $v_{j,4}^{(L)} = v_{j,4}^{(H)}$. 

The Hamiltonian in Eq.~(\ref{eq:Htot1}) can be re-expressed in terms of the system-plus-bath model
\begin{align}
\label{eq:Htot2}
\hat{H} _{tot} =\hat{H} _{s}+ \hat{H} _{b}+\hat{H}_{sb}.
\end{align}
Here we set the Planck constant $\hbar$ and  the Boltzmann constant $k_{B}$ to unity. 
The Hamiltonian for a two-state system in Eq.~(\ref{eq:Htot2}) can be represented 
as $\hat H_{s}=-\frac{\Delta}{2}{\sigma }_{z} -\frac{\lambda}{2}{\sigma }_{x}$, 
where $\Delta=\epsilon_H - \epsilon_L$ is the energy bias between the LS and the HS 
states and ${\sigma_z}/{\sigma_x}$ 
are the spin-1/2 Pauli matrices. Meanwhile, the Hamiltonian for the thermal bath is given 
by $\hat{H}_{b} =\sum_{j}[\hat{p}_{j}^{2} /(2m_{j})+V_{j}(\hat{x}_{j})]$, 
where $V_{j}(\hat{x}_{j})=\frac{1}{2}m_{j} \omega _{j}^{2}\hat{x}_{j}^{2}
(1+b_{j}\sqrt{\omega_{j}}\hat{x}_{j}+a_{j}\omega_{j}\hat{x}_{j}^{2})$ 
is the potential of the thermal bath, with $a_{j}$ and $b_{j}$ being coefficients 
characterizing bath anharmonicity. The Hamiltonian for the system-bath interaction 
is $\hat{H}_{sb}=\sigma _{z}\sum_{j}c_{j}(\hat{o}_{j}-\langle\hat{o}_j\rangle)$, 
where $c_{j}$ denotes the coupling constant between the system and the environment, 
the operator $\hat{o}_{j}= \hat{x}_{j}+\kappa_{j} \sqrt{\omega _{j}} \hat{x} _{j}^{2}$ 
with $\kappa_{j} $ being the nonlinear strength in the spin-vibration coupling, 
and $\langle\hat{o}_j\rangle=\textrm{Tr}[\hat{o}_je^{-\beta\hat{H}_b}]/\textrm{Tr}e^{-\beta\hat{H}_b}$ 
denotes the equilibrium expectation of the operator $\hat{o}_j$.

Utilizing the linear response approximation, the effect of the bath is encapsulated by its
correlation function~\cite{ilk1994real,makri1999linear,makri1999iterative}
\begin{align}
\alpha(t)=\frac{1}{\pi } \int_{0}^{\infty }\mathrm{d}\omega J_{\beta }(\omega)[\coth (\frac{\beta \omega}{2})\cos \omega t-i\sin \omega t],
\end{align}
where $\beta = {1}/{T}$ with $T$ being the temperature, and $J_{\beta }(\omega)$ is the effective spectral density function
\begin{align}
\label{eq:Htot4}
J_{\beta }(\omega )=\sum_{j,m,n}\frac{c_{j}^{2}\pi}{4Z_{j}}(e^{-\beta \epsilon_{n}^{(j)}}-e^{-\beta \epsilon_{m}^{(j)}})|o_{m,n}^{(j)}|^{2}\delta (\omega-\omega_{mn}^{(j)} ).
\end{align}
Here $j$ is the index of the bath mode, $\epsilon_{n}^{(j)}$ denotes the ${n}$th eigen-energy, 
$Z_{j}$ represents the partition function, and $\omega_{mn}^{(j)}$ stands for the transition frequency for $m\to n$. 
In the weak spin-vibration coupling regime under investigation, the  Redfield equation can be used to obtain the equilibrium expectation
\begin{align}
  \label{eq:qme}
\frac{\mathrm{d} \hat{\rho}(t)}{\mathrm{d}t}={i}[\hat{\rho}(t),\hat{H}_{s}] -[{\sigma}_z,(\hat{\Xi}\hat{\rho}(t)-\hat{\rho}(t)\hat{\Xi}^{\dagger})],
\end{align}
where the operator $\hat{\Xi}$ is defined as
\begin{align}
\hat{\Xi}=\int_{0}^{\infty} \mathrm{d}\tau \alpha(\tau)e^{-{i}\hat{H}_{s}\tau} {\sigma}_z e^{{i}\hat{H}_{s}\tau} .
\end{align}

Equation~(\ref{eq:qme}) is the working equation of this work. Some remarks are in order. First,
Eq.~(\ref{eq:qme}) and the underlying system-bath paradigm 
are based on the assumption of a weak spin-vibration coupling. 
A natural question arises: How weak is weak enough to be handled with the above scheme?
The answer depends on specific systems. Here we discuss this issue with a specific setting mimicking 
the typical temperature-induced SCO, {\it i.e.}, $\Delta = \lambda=50$ meV, $T<300$ K, and with a 
high-frequency cutoff of 800 cm$^{-1}$ for the heat bath.
Analog to the linear dissipation, we adopt the renormalization energy $E_r = \int^\infty_0 d\omega J_\beta(\omega)/\omega$ 
to characterize the overall coupling strength. Model simulations show that in the absence of anharmonicity, 
calculations with $E_r = 25$ meV can produce reliable results. 
In the presence of significant anharmonicity, we should be more cautious and limit the dissipation strength 
up to 10 meV.

In realistic SCO systems the spin-vibration coupling is not necessarily weak. 
For instance, in the two-dimensional layer 
[Fe$^\textrm{II}$((3,5-(CH$_3$)$_2$Pz)$_3$BH)$_2$] (Pz = pyrazolyl), 
the spin-vibration coupling constants of one or two modes lie between 20 meV 
and 50 meV and the rest are below 10 meV~\cite{bairagi16_12212}.
Such a separation that the spin-vibration coupling is dominated by one or 
two particular modes is not unique for [Fe$^\textrm{II}$((3,5-(CH$_3$)$_2$Pz)$_3$BH)$_2$] 
but widely exists in many SCO complexes~\cite{lemke17_15342,svensson24_1935}.
In this case we can follow the treatments adopted in the context of exciton dynamics 
to include these modes into the system
and treat the rest modes as a heat bath~\cite{kuehn96_99}. The remaining 
spin-vibration coupling then becomes sufficiently weak to allow a QME treatment.
The above procedure is therefore applicable with a direct enlargement of the system.

\begin{figure}[t!]
\centering
\includegraphics[scale=0.33]{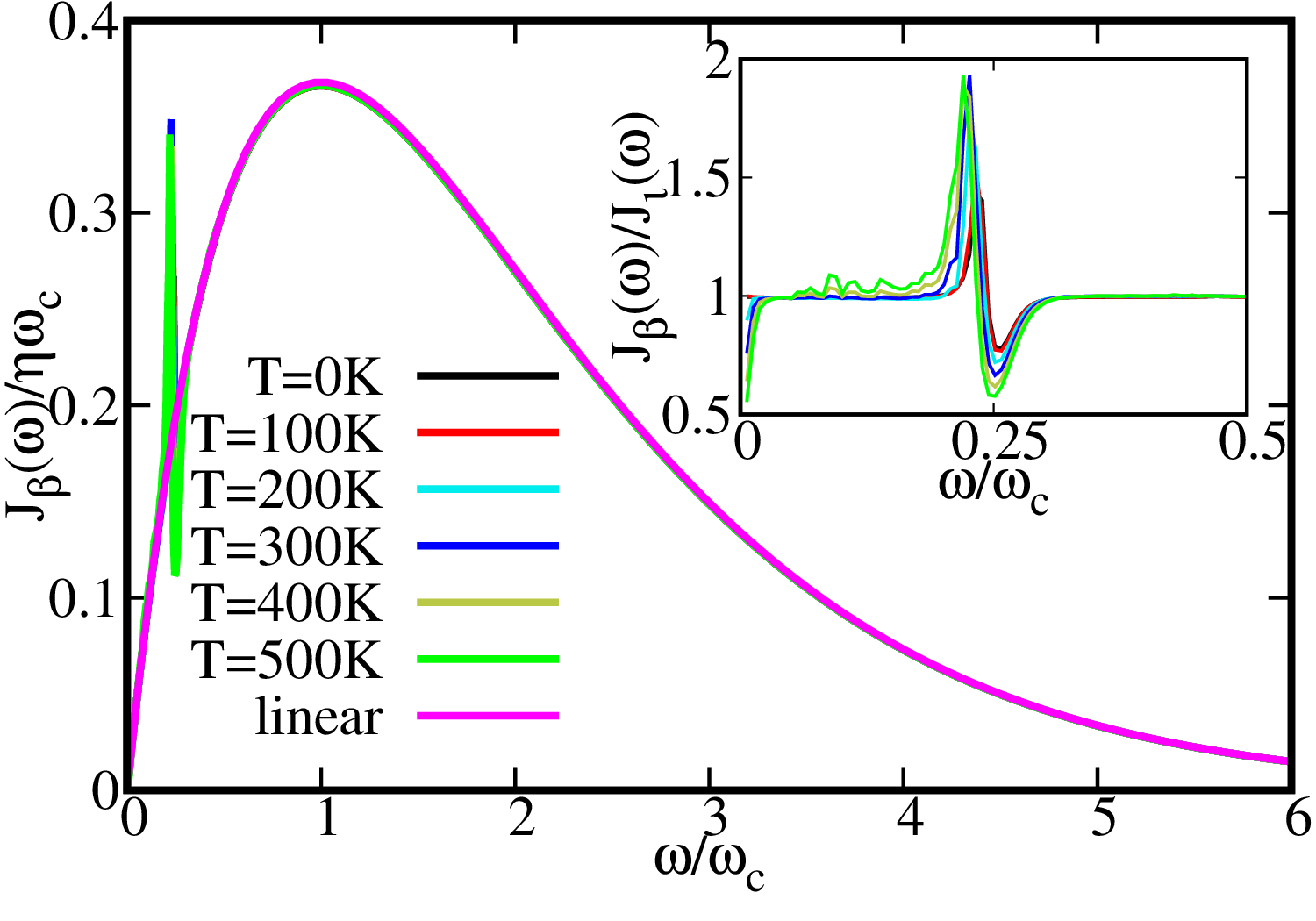}
\caption{Effective spectral density function at different temperatures.  The parameters 
 in the nonlinear dissipation part are 
$a_j=0.0043253\omega_j^3/((15 + \omega_j)(1 - 31.52 \omega_j^2 + 256 \omega_j^4)^2)$,
$b_j=-0.49020\omega_j^{3/2}/((15 + \omega_j)(1 - 31.52\omega_j^2 + 256\omega_j^4))$, 
and $\kappa_j=0$. 
The parameters $c_j$ and $\omega_j$ for the linear part is determined by discretizing  
the spectral density $J_{\iota} (\omega)=0.05 \omega \exp(-\omega/\omega_c)$ with 20 000 modes. Check text for details about the Hamiltonian.\label{fig:1}}
\end{figure}

Second, beyond the weak coupling regime, a universal nonlinear-dissipation theory is 
yet to be developed but methods are available for specific cases.
For systems that can be reasonably modeled with
the lowest-level nonlinearity (harmonic potentials with  
linear plus quadratic spin-vibration couplings), 
more advanced methods, such as the hierarchical equation of motion~\cite{xu18_114103} 
and the quantum stochastic Liouville equation~\cite{yan19_074106} are useful to tackle the physics.

Third and the last, here we focus on transitions in a single molecule and therefore not include
cooperative interactions between metal centers which are pivotal to implement practical SCO materials. 
However, the open system paradigm, as a generic framework to tackle quantum dissipative dynamics, 
can be straightforwardly extended to oligomers or lattices. To be specific, the Hamiltonian $\hat{H}_{tot}$ 
can be generalized to a nonlinearly-dissipated quantum Ising model~\cite{jin18_241108,hur18_451}, that is, 
$\hat{H}_s = \sum_a \left[\frac{\Delta_a}{2}\sigma_a^x\right. + \left.\frac{\epsilon_a}{2}\sigma_a^z\right]
   + \sum_{ab} K_{ab} \sigma_a^z \sigma_b^z$ and 
   $\hat{H}_{sb} = \sum_a \sigma_a^z\sum_j c_{j,a}(\hat{o}_{j}-\langle\hat{o}_j\rangle)$.
Here $\sigma_a^z$($\sigma_a^x$), $\Delta_a$, and $\epsilon_a$
denote the Pauli matrices, the energy bias, and the spin-orbital coupling 
on site $a$, respectively, 
$c_{j,a}$ stands for the coupling constant between 
the $a$th spin  and the $j$th vibration,
and $K_{ab}$ refers to the nearest neighbor interaction along $z$-direction. 
In this model the cooperative effects are encoded in the effective direct spin-spin interaction $K_{ab}$ 
and the coupling of different spins to the same vibrational modes. 
As illustrated by Wolny and coworkers~\cite{rackwitz13_15450,wolny16_19},
the parameters can again be extracted from {\it ab initio} calculations.

\begin{figure}[t!]
\centering
\includegraphics[scale=0.35]{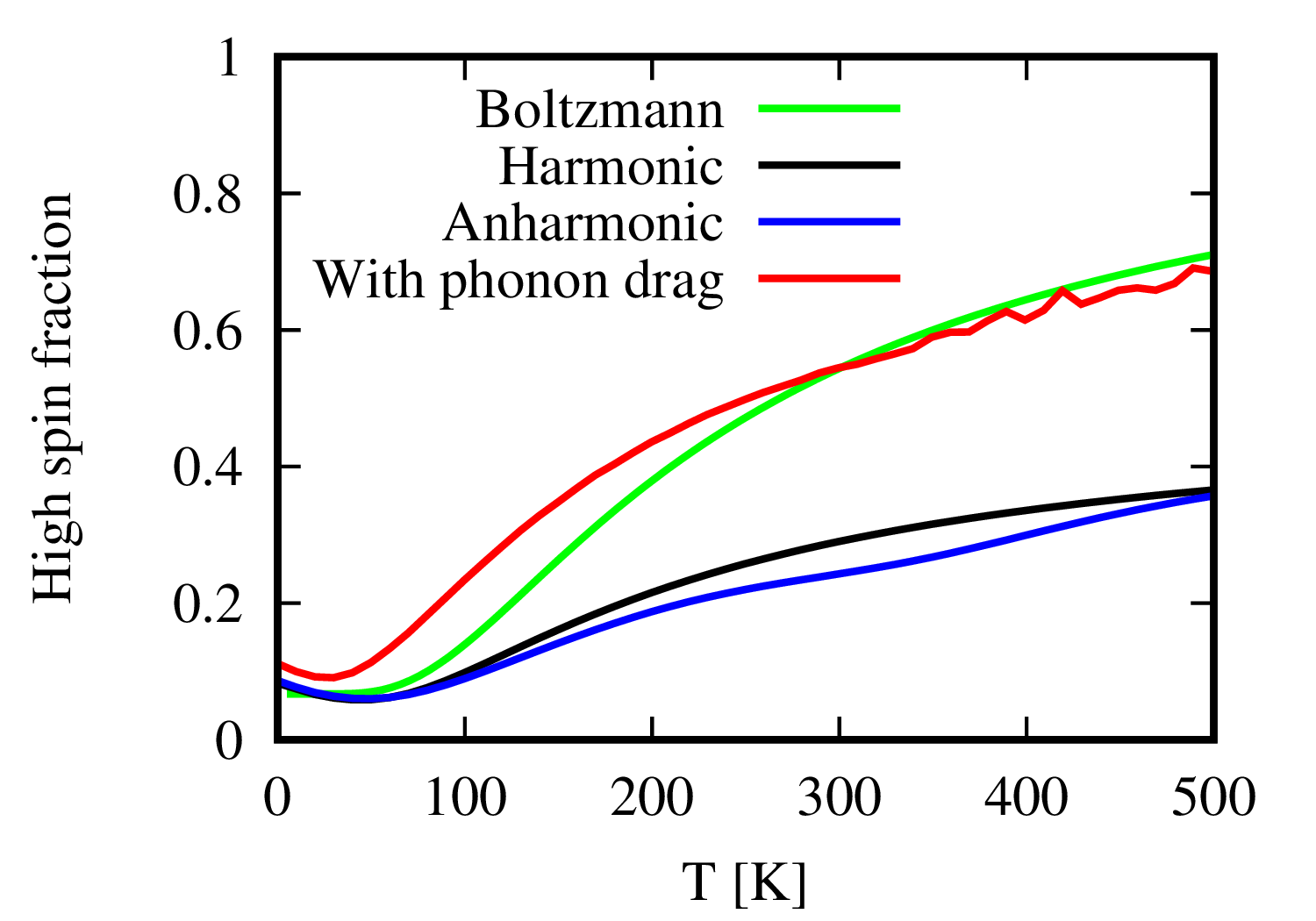}
\caption{The effects of the anharmonicity in the bath potential on SCO.  
Boltzmann: Results given by the Boltzmann distribution; 
Harmonic: Equilibrium results of the Redfield equation with $a_j=0$, $b_j=0$, and $\kappa_j=0$; 
Anharmonic: Equilibrium results with anharmonicity in the bath potential only; 
With phonon drag: The thermal average $\langle \hat{o}_j\rangle$ is included in energy splitting. 
Parameters $\kappa_j$, $a_j$, and $b_j$  for the latter two are the same as that in Fig.~\ref{fig:1}.\label{fig:2}}
\end{figure}

\section{Numerical results and discussions}
Here we adopt a discretized description for the anharmonic bath~\cite{wang2007quantum}.
To this end, we determine the parameters $c_j$ and $\omega_j$ upon 
discretizing the spectral density function $J_{\iota}({\omega})=\eta \omega \exp ({-\omega /\omega _{c}})$ with 20 000 modes, 
where $\eta$ is the linear dissipation strength and $\omega_c$ is the high frequency cutoff.  
For the anharmonic coefficients, we set 
$a_j=\omega_{c}/({2\omega_{j} x_{j,-} x_{j,+}})$, 
$b_j=-2\sqrt{\omega_{c}}(x_{j,-} + x_{j,+})/(3\sqrt{\omega_{j}}x_{j,-}x_{j,+})$, 
where $x_{j,\pm }=f(\omega_{j}/\omega_{c})(1\pm 0.2\sqrt{1- 1.6 \omega_{j}/{\omega_{c}}})$ with 
$f(u)=\alpha(8u^{2}+ 1/32u^{2}+\gamma)$. Here  $\alpha$ and $\gamma$ are two parameters controlling anharmonicity.  
Under these conditions, the bath modes with $\omega_j<\omega_c/1.6$ assume  a double-well potential. 

In this study we set $\Delta=300$ K,  $\lambda=50$ meV, $\eta=0.05$, $\omega_ {c}=800$ cm$^{-1} $, $\alpha=1360 $, 
and $\gamma=-0.985$. The parameters $\kappa_j$ will be set to the same value for all vibrations, varying from 0 to 0.05 in step of 0.01. 
The calculated correlation functions for  temperatures  
from 0 K to 500 K with intervals of 10 K are substituted into the Redfield equation to obtain the equilibrium distributions. 

In Fig.~\ref{fig:1} we present the effective spectral density functions for temperatures from 0 K to 500 K 
in step of 100 K. In the simulations, we first calculate the bath correlation function and perform 
the Fourier transform of its imaginary part to obtain the effective spectral density functions. 
For comparison we also plot the ratio ${J_{\beta }(\omega)}/{J_{\iota} (\omega)}$. As illustrated in Fig.~\ref{fig:1}, 
we observe that the effective spectral density function becomes temperature-dependent
 and deviates from the linear one.  The overwhelming feature is that a sharp peak appears
 around $\omega=0.22 \omega_{c}$ besides the original peak of the linear spectral density function. 
The anharmonic results merge to linear dissipation when $\omega > 0.4\omega_{c}$.

We now discuss the effect of the anharmonicity in the bath potentials without the nonlinearity in the spin-vibration coupling.
The results with linear dissipation are also presented. As shown in Fig.~\ref{fig:2}, with the anharmonic model, 
the HS fraction first decreases and then gradually increases with temperature. Even at a temperature as high as 500 K, 
the HS fraction is still less than one half. The harmonic model assumes roughly the same trend as the anharmonic one.
 However, subtle differences exist between the two trends, mainly in the temperature range from 100 K to 300 K, 
which is exactly the region in which the effective spectral density function deviates from its linear counterpart.
It seems that the presence of the anharmonicity in the bath potential alone slows down the LS to the HS transition.
For comparison we show the results obtained from the Boltzmann distribution. It is interesting to note that even for such a weak dissipation,
 the results from QME are significantly different from the Boltzmann distribution.

\begin{figure}[t!]
\centering
\includegraphics[scale=0.35]{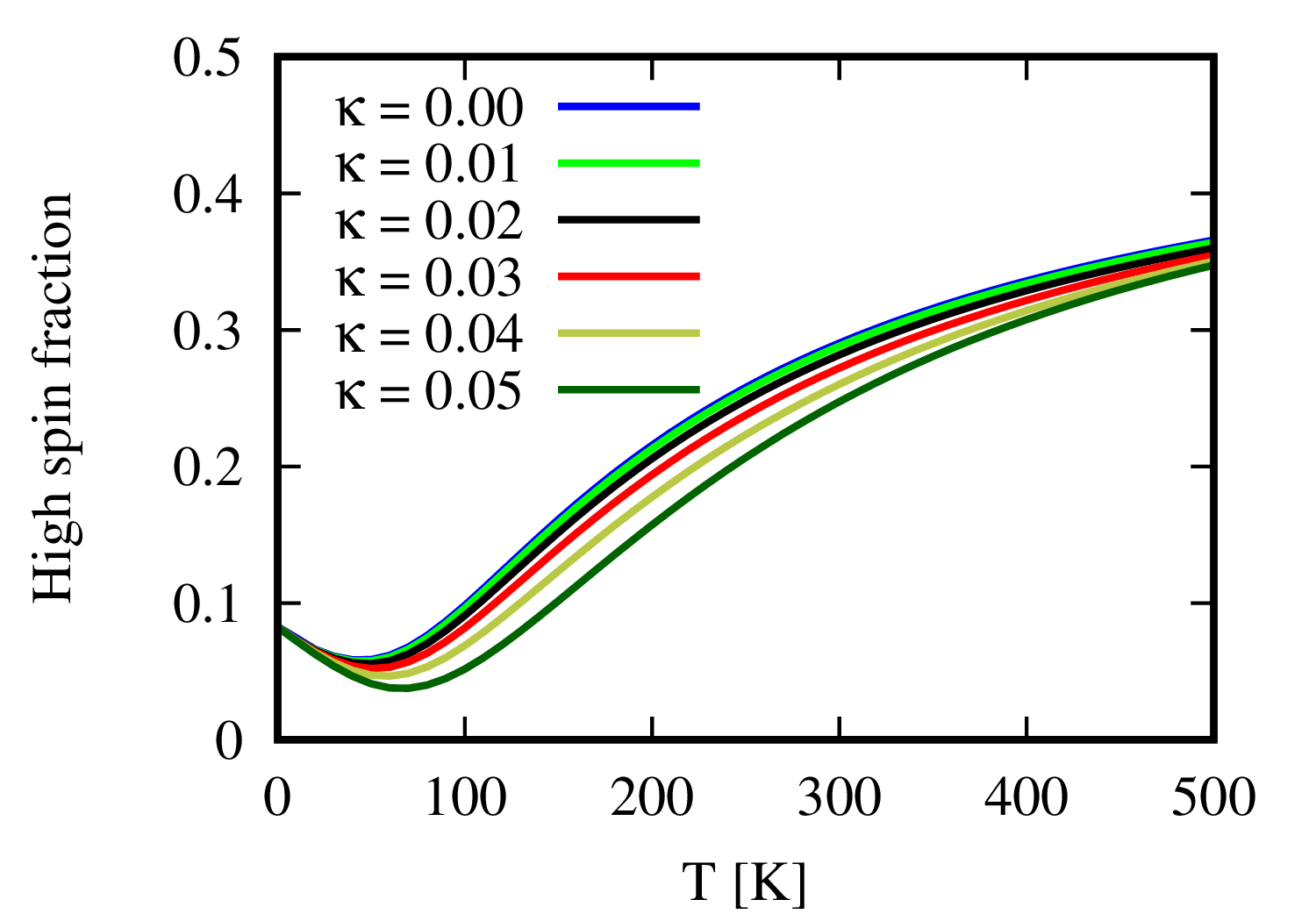}
\caption{The effects of nonlinear spin-vibration coupling with harmonic bath potential. 
The parameter $\kappa_j$ are the same for all vibrations.\label{fig:3}}
\end{figure}

Next we investigate the effect of the nonlinearity in the spin-vibration coupling without the potential anharmonicity. 
The results are shown in Fig.~\ref{fig:3}, which depicts the same overall trend as that in Fig.~\ref{fig:2}. 
For different nonlinear intensities characterized with $\kappa$, the differences are pretty small and only obvious between 100 K and 300 K. 

\begin{figure}[t!]
\centering
\includegraphics[scale=0.35]{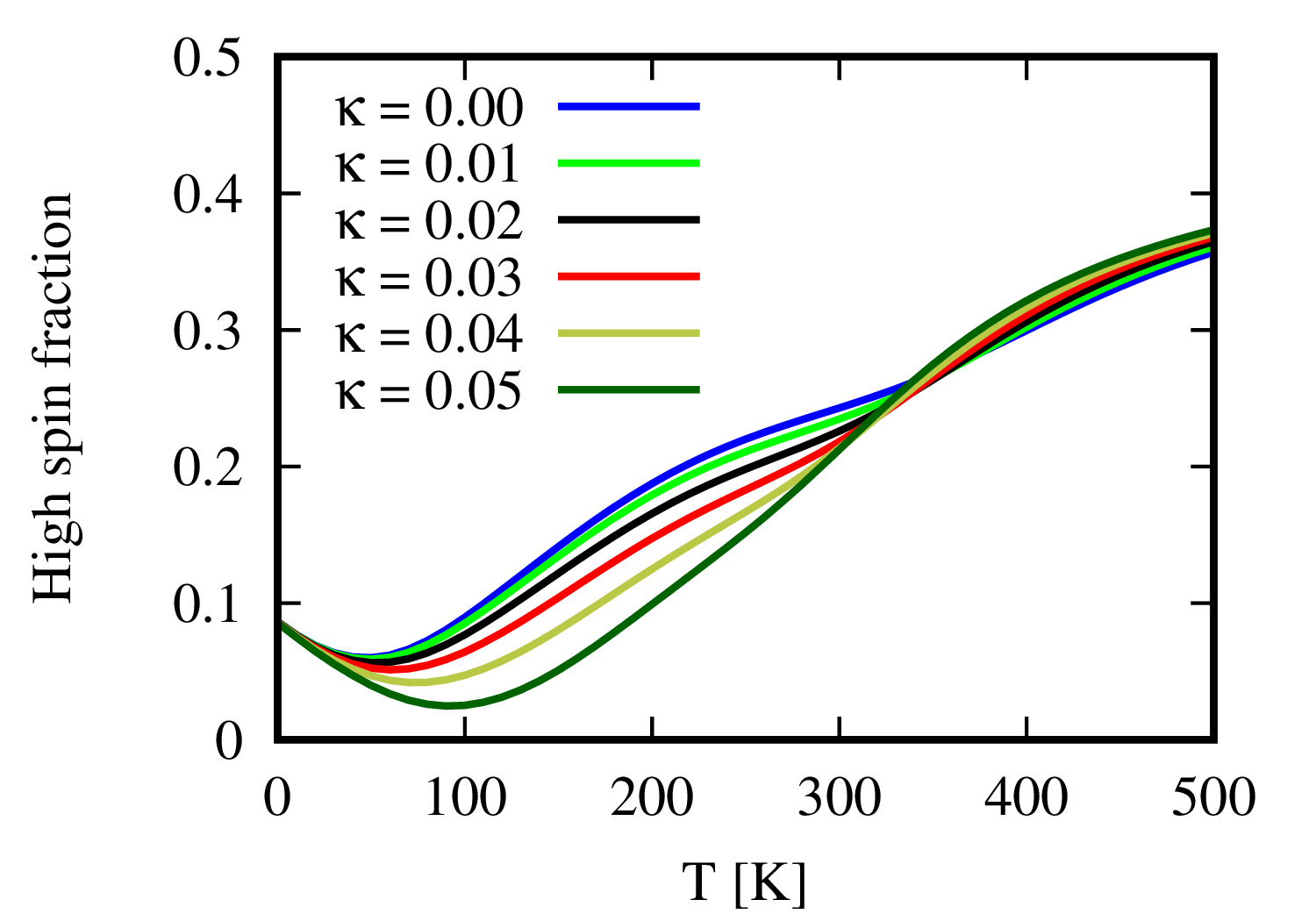}
\caption{The interplay of the nonlinearity in the spin-vibration coupling and the anharmonicity in the potential.  
Anharmonic parameters $a_j$ and $b_j$ are the same as that in Fig.~\ref{fig:1}.\label{fig:4}}
\end{figure}

In the presence of both the anharmonicity in the bath potential and the nonlinearity in spin-vibration coupling, 
there may have strong interplay between these two effects. As shown in Fig.~\ref{fig:4}, now the temperature dependence 
are much more significant than that in Figs.~\ref{fig:2} and \ref{fig:3}, 
especially in the temperature range from 100 K to 300 K. With such an interplay, 
the HS fraction changes much abrupter with temperature 
for stronger nonlinear spin-vibration coupling.

The vibrational anharmonicity will affect SCO via another mechanism. When the system jumps between 
the HS and the LS states, the vibrational environment the spin feels 
is switched between $\hat{H}^{(L)}_{vib}$ and $\hat{H}^{(H)}_{vib}$ that are specified in Eq.~(\ref{eq:Htot1}).
From the view point of reduced dynamics,
the energy splitting between HS and LS is in principle
$\Delta = \epsilon_H-\epsilon_L  + \langle\hat{H}^{(H)}_{vib}-\hat{H}^{(L)}_{vib}\rangle$.  
The thermal equilibrium expectations $\langle\hat{H}^{(S)}_{vib}\rangle$ ($S$ = $H$, $L$) 
differ for different spin states in the presence of potential anharmonicity,
and the molecular structure will be distorted according to the spin change. 
This equilibrium shift, which is absent for harmonic baths,  
introduces an additional energy splitting $\sum_j c_j \langle\hat{o}_j \rangle$ to SCO.
Such a back-action mechanism of spin-induced molecular structure distortion 
is essentially the phonon drag effect observed in 
thermoelectric transportation~\cite{frederikse53_248,herring54_1163,gurevich89_327}.
However, there are subtle differences between these two cases. 
In thermoelectric transportation the phonon drag stems from the phonon motion against the temperature gradient 
and is mostly evidenced at low temperatures.
In temperature-induced SCO the key role of drag effects is due to 
the temperature-dependent energy splitting reflecting the
temperature-dependent distortion of the molecular structure. 
It is expected that the drag effects in SCO are manifested 
for temperatures under which the average molecular structures are significantly different 
between the HS and the LS configuration.
To characterize this phonon drag effect, we set all $c_j$ coefficients 
to positive in the bath discretization and add the thermal average $\sum_j c_j \langle\hat{o}_j \rangle$ to $\Delta$. 
Note that in calculating the thermal average $\langle \hat{o}_j\rangle$, 
the quadratic term $\hat{x}^2_j$ always assumes a temperature-dependent result. 
In order to simplify the analysis and focus on the phonon drag effect we 
use $\kappa=0$ to exclude the effect of nonlinear spin-vibration coupling.  
The corresponding results are demonstrated in Fig.~\ref{fig:2}, which shows that the phonon drag 
leads to a more pronounced and much abrupter temperature-dependence of the HS fraction.

\section{Summary and outlook}

In metal-organic compounds, vibrations may assume strong anharmonicity.
To take account of the anharmonicity effect on the temperature-induced SCO, 
a linear dissipation model has to adopt a separate spectral density function 
at each temperature and thus fails to describe the temperature-dependent transition.
Here an anharmonic vibrational bath model is used to simulate single-molecule SCO. 
With the linear response approximation, we are able to obtain the effective spectral 
density functions for all temperatures consistently with the same bath that can be extracted from 
{\it ab initio} calculations.
To scrutinize the anharmonicity effect itself, we focus on the weak spin-vibration coupling to 
avoid further complications caused by strong interaction. 
Propagating the Redfield equation to sufficiently long time, we can obtain the 
equilibrium distribution of the spin. 

With specific double-well potentials for low-frequency vibrations, we have shown that 
the effective spectral density functions assume significant temperature dependency. 
We have performed four series of calculations with the obtained 
temperature-dependent spectral density functions: 
(1) anharmonicity in the bath potentials only;
(2) nonlinearity in spin-vibration couplings only;
(3) anharmonic bath potentials together with nonlinear spin-vibration couplings;
(4) including the energy difference associated with the spin-induced molecular structure distortion.
We have revealed that nonlinearities in the couplings or the potentials alone 
produce weak effects but their combination yields much stronger influence.
Further, we have demonstrated that in the presence of anharmonicity, 
the SCO is drastically affected by the spin- and temperature-dependent thermal-average of 
vibrational degrees of freedom and becomes much abrupter. 
We have called it the phonon drag effect because 
it is essentially the same mechanism first found in thermoelectric transportation.

Here we only consider single-molecule transitions in the weak spin-vibration coupling regime. 
With the nonlinearly-dissipated quantum Ising model, our approach can be extended to study 
the cooperative effects in a molecular chain or lattice.
Further, we can go beyond weak coupling assumption for those systems in which the spin-vibration 
interaction is dominated by one particular mode.
Such a mode can be approximately included in QME 
as a two-level system because it is generally a breathing vibration and 
its second- and higher-excited states are only slightly populated at temperatures under 300 K. 
Then, with today's moderate computational resources, 
we can simulate the physics of a system up to 8 sites if this particular vibration is 
explicitly treated with QME and up to 16 sites otherwise. As such, we can have 
a consistent quantum description to scrutinize the interplay between anharmonicity, 
center-center cooperative interactions, and other factors.

\section*{Acknowledgments}

The authors acknowledge the support from the National
Natural Science Foundation of China under grant No. 21973036.


%
\end{document}